\documentclass[hyperref,UTF8]{ctexart}  

\usepackage{amsmath}         
\usepackage{graphicx}        
\usepackage{float}
\usepackage{varwidth}
\usepackage{hyperref}       
\usepackage{geometry}        
\usepackage[format=hang,font=small,textfont=it]{caption}
\usepackage[nottoc]{tocbibind}   
\usepackage[greek,english]{babel}
\usepackage{pstricks,pst-plot,pst-grad,pst-3dplot,pstricks-add,pst-node} 

\newcommand{\beq}[1]{\begin{eqnarray}\label{#1}}
\newcommand{\eeq}{\end{eqnarray}}

\title{Variation of entanglement entropy and mutual information in fermion-fermion scattering  }
\author{Jinbo Fan, Yanbin Deng, Yong-Chang Huang
\\
\\\textit{Institute of Theoretical Physics, Beijing University of Technology}
\\ \textit{Beijing, 100124, China} 
\\
\\ ~~~~~~~~~~~~~\textit{Email:} fanjinbo@emails.bjut.edu.cn
\\ ~~~~~~~~~~\textit{Email:} scientifichina@outlook.com
\\ \textit{Email:} ychuang@bjut.edu.cn}
\date{\today}
\begin{document}

\maketitle                      % 命令实际输出论文标题
\begin{abstract}
We study the behavior of entanglement between different degrees of freedom of scattering fermions, based on an exemplary QED scattering process $e^+e^-\longrightarrow\mu^+\mu^-$.  The variation of entanglement entropy between two fermions from an initial state to the final state was computed, with respect to different entanglement between the ingoing particles. This variation of entanglement entropy is found to be proportional to an area quantity, the total cross section. We also study the spin-momentum and helicity-momentum entanglements within one particle in the aforementioned scattering process. The calculations of the relevant variations of mutual information in the same inertial frame reveals that, for a maximally entangled initial state, the scattering between the particles does not affect the degree of both of these entanglements of one  particle in the final state. It is also found that the increasing degree of entanglement between two ingoing particles would restrict the generation of entanglement between spin (helicity) and momentum of one outgoing particle. And the entanglement between spin and momentum within one particle in the final state is shown to always be stronger than that for helicity-momentum for a general initial entanglement state,  implying significantly distinct properties of entanglement for the helicity and spin perceived by an inertial observer.
\end{abstract}
  
\newpage
\tableofcontents  

\section{Introduction} 

 The physics of fermion-fermion scattering plays a crucial role in a wide variety of scattering experiments, which probe the behavior of elementary particles. Nevertheless, these theoretical investigations often focus on classical observables such as cross section and decay rate, since these quantities are simpler to access via experiment. An essential property that distinguishes quantum mechanics from classical mechanics is the possibility of entanglement between different degrees of freedom. It is a characteristic feature of some states of a composite system that cannot be decomposed into a direct product form of subsystems. These states also called  entanglement states. Entanglement entropy is a measure of how much a given quantum state is quantum mechanically entangled.  In the paper, we investigate issues of entanglement entropy in simple example of fermion-fermion scattering.

 Even though quantum information was originally formulated in terms of nonrelativistic quantum mechanics, recent years have seen increasing research interest in studying it within the more fundamental framework of quantum field theory. Calabrese and Cardy have systematically studied issues of entanglement entropy in the quantum field with the use of a replica trick \cite{1,2}. The theoretical framework to study entanglement entropy in cosmology  is the quantum field theory in curved background \cite{3,4,5}. An expanding spacetime generates entanglement between certain modes of an exclusively gravitationally interacting scalar field, whose entanglement entropy contains information about the parameters of the cosmic history \cite{6,7}. When one considers a composite system with subsystem A and its complement $\bar{A}$ separated by a surface, the entanglement entropy is proportional to the area of the surface, and depends on the UV cutoff, which regulates the short-distance correlations. Applied to black hole, Solodukhin \cite{9} calculates the entanglement entropy  when the entangling surface is the black-hole horizon. In  Refs.\cite{8,9}, the entanglement entropy can be interpreted not as the total but as a partial (quantum corrections) contribution to the black hole entropy. A more complete understanding may arise from the AdS/CFT.
 
   Ryu and Takayanagi \cite{10,11} achieved the holographic derivation of entanglement entropy in quantum (conformal) field theories from the perspective of AdS/CFT. In these articles, the entanglement entropy between A and $\bar{A}$ was obtained by evaluating $S_E=\mathcal{A}/(4G_N)$, where $\mathcal{A}$ is the area of a minimal surface whose boundary is the boundary of the subsystem A. These provide a geometric understanding of entanglement. Furthermore, the ER=EPR conjecture by Maldacena and Susskind \cite{12} provides us with another geometric interpretation for entanglement entropy. The two distant black holes connected through a Einstein-Rosen bridge (or a wormhole) in the interior  can be interpreted as corresponding to a maximally entangled state of two black holes that form a complex Einstein-Rosen-Podolski pair. They suggest that entangled states contain similar bridges in general. The entanglement between two particles, which are, for example, a pair of accelerating quark and antiquark, the EPR pair created via the Schwinger effect and a pair of scattering gluons in strongly coupled super Yang-Mills theory (SYM), had been studied [13,14,15,16]. These research attempts gave some supportive examples for the ER=EPR conjecture.
    
    	In the AdS/CFT correspondence, the scattering amplitude in a strongly coupled field theory can be related to the area of the minimal surface of the Wilson loop of trajectories of scattering particles \cite{17,18}. On the other hand, the holographic entanglement entropy is proportional to the area of a minimal surface in AdS spacetimes. In a word, both the scattering amplitude and entanglement entropy in a strongly coupled field theory are associated with minimal surfaces from the point of view of the AdS/CFT correspondence. References.\cite{19,20} studied the entanglement entropy of two divided momentum spaces with the perturbative calculations method, and this method was then followed by Refs.\cite{21,22} for the study of the entanglement between two scalar particles in the scattering process in a weakly coupled field theory. They found that the entanglement entropy changes during the scattering process, and this variation of entanglement entropy from initial state to final state is proportional to the cross section. 
    	
    	Being attracted by the interestingness of the previous research, several authors invested multiple attempts to study the behavior of the entanglement between particles in the scattering process from different perspectives \cite{23-25}. But all these efforts, focusing on elastic scattering of scalar particles and only considering the entanglement of momentum degrees of freedom, worked in toy models. All matter particles and antimatter particles in nature are the fundamental fermions (quarks, leptons, antiquarks, and antileptons). In addition to the momentum degree of freedom, fermions carry half-integer spin and helicity degrees of freedom not possessed by scalar particles. It is worthwhile to mention that in quantum information processing, the spin of a particle is often used as a qubit regardless of the momentum state of the particle. However, spin and momenta are not separable in general in the relativistic motion. To better generalize the entanglement behaviors of scattering scalar particles to the general fields, we need to study it in a fundamental model. The goal of this paper is to carry out the perturbative calculations method in \cite{21,22} for fermion-fermion scattering, i.e. the simplest QED process $e^+e^-\longrightarrow\mu^+\mu^-$.

	In Sec. 2, we give the variation of entanglement entropy of the two scattering fermions with respect to different entanglement of the initial state, and we consider the QED scattering process $e^+e^-\longrightarrow\mu^+\mu^-$ as an example of a case study.  The spin state and the helicity state are the basis of Hilbert space for the Dirac field, however, spin entanglement and helicity entanglement are hinted of different properties by Ref. \cite{26,27}. In Sec. 3, we numerically analyze the mutual information between spin (helicity) and momentum degrees of freedom with one fermion in scattering process. Section 4 is devoted to the conclusion and discussion.

\section{Entanglement entropy in fermion-fermion scattering}

\subsection{Entanglement entropy}
We consider the scattering process of two fermionic fields, $\psi_A, \psi_B$, with the Hamiltonian    $H=H_{\textup{free}}+H_{\textup{int}}$. For an elastic scattering process of two fermions, the Hilbert space for both the initial state and final state would be (1+1)-particle Fock space. At weak coupling, we can assume the unitarity of local interaction terms to be guaranteed at lower orders of perturbation \cite{21}. The initial and final states  can be viewed as superposition of the basis of free Hamiltonian $H_{\textup{free}}$ so that we can divide the total Hilbert space as $\mathcal{H}_{\textup{tot}}=\mathcal{H}_A\otimes\mathcal{H}_B$.

Since incoming and outgoing particles are free on-shell particles, we can describe the (1+1)-particle states as
\begin{align}
\vert p,s;q,r\rangle=\sqrt{2E_{\textbf{p}}}~{a^s_{\textbf{p}}}^{\dagger}\vert0\rangle_A\otimes \sqrt{2E_{\textbf{q}}}~{b^r_{\textbf{q}}}^{\dagger}\vert0\rangle_B
\end{align}
where $\textbf{p}$ and $\textbf{q}$ are the 3-momenta of particles, and $s$, $r$ denote the spin or helicity of the particle.
The fermionic creation/annihilation operators obey the commutation relations,
\begin{align}
\{a^s_{\textbf{p}}, {a^r_{\textbf{k}}}^\dagger\}=(2\pi)^3\delta^{(3)}(\textbf{p}-\textbf{k})\delta^{sr}
,~~~~
\{b^n_{\textbf{q}}, {b^m_{\textbf{\textit{l}}}}^\dagger\}=(2\pi)^3\delta^{(3)}(\textbf{q}-\textbf{\textit{l}})\delta^{nm}
\end{align}
and the inner product between 2-particle states is defined as
\begin{align}
\langle k,s^\prime; l,r^\prime\vert p,s;q,r\rangle=2E_{\textbf{k}}2E_{\textbf{\textit{l}}}(2\pi)^3\delta^{(3)}(\textbf{k}-\textbf{p})  (2\pi)^3\delta^{(3)}(\textbf{\textit{l}}-\textbf{q})\delta^{ss^\prime}\delta^{rr^\prime}
\end{align}
These creation/annihilation operators are the mode coefficient of Fourier expansion of free fermion fields
\begin{align}
\bar{\psi}&=\int\frac{d^3\textbf{p}}{(2\pi)^3}\frac{1}{\sqrt{2E_{\textbf{p}}}}\sum_{r=1,2} \left(b^r_{\textbf{p}}~\bar{\nu}^r(p)e^{-ip\cdot x}+{a^r_{\textbf{p}}}^{\dagger}~\bar{\mu}^r(p)e^{ip\cdot x}
 \right)
\\ \notag
\psi&=\int\frac{d^3\textbf{q}}{(2\pi)^3}\frac{1}{\sqrt{2E_{\textbf{q}}}}\sum_{s=1,2} \left(a^s_{\textbf{q}}~\mu^s(q)e^{-iq\cdot x}+{b^s_{\textbf{q}}}^{\dagger}~\nu^s(q)e^{iq\cdot x} \right)
\end{align}
For a scattering process, the final state is determined by the initial state and the $S$ matrix\cite{21},
\begin{align}
\vert \textup{fin}\rangle=\int\frac{d^3\textbf{k}}{(2\pi)^3}\frac{1}{2E_{\textbf{k}}}\frac{d^3\textbf{\textit{l}}}{(2\pi)^3}\frac{1}{2E_{\textbf{\textit{l}}}}\sum_{s^\prime,r^\prime}\vert k,s^\prime;l,r^\prime\rangle\langle k,s^\prime;l,r^\prime\vert\textbf{S}\vert p,s;q,r\rangle
\end{align}
The $T$ matrix be defined as 
\begin{align}
\textbf{S}=\textbf{1}+i\textbf{T}
\end{align}
and the invariant matrix element $\mathcal{M}$,
\begin{align}
\langle k,s^\prime;l,r^\prime\vert i\textbf{T}\vert p,s;q,r\rangle=(2\pi)^4\delta^{(4)}(p+q-k-l)\times i\mathcal{M}
\end{align}

The authors of Ref.\cite{21} and Ref.\cite{23} analyzed the entanglement between scalar particles, where the entanglement occurs  among momentum degrees of freedom. What is more interesting is the  entanglement between the spin degrees of freedom for fermionic particles. In the following, we will give the entanglement entropy between fermionic degrees of freedom. The process of evaluation  will be the following: $\vert \Psi\rangle\rightarrow\rho_{AB}\rightarrow\rho_A\rightarrow S_E$.
                                                                                                                                                                                                                                                                                                                                                                                                                                                                                                                                                                                      
 We choose spin states as the basis of Hilbert space, and consider the following initial state with parametrization of the entanglement between the spins:
\begin{align}
\label{ini}
\vert\textup{ini}\rangle=\cos\eta\vert p,\uparrow; q,\uparrow\rangle+\sin\eta~ e^{i\beta}\vert p,\downarrow; q,\downarrow\rangle
\end{align}
with $(\uparrow,\downarrow)\equiv\sigma$ representing the spin along the $z$ axis, $\eta\in[0,\pi/2]$ parametrizing the spin entanglement of the state, and $\beta\in[-\pi/2,3\pi/2]$ labelling the relative phase of the superposed states $\vert p,\uparrow; q,\uparrow\rangle$ and $\vert p,\downarrow; q,\downarrow\rangle$. For $\eta=0$, or $\eta=\pi/2$, the initial state is not an entangled state. For $\eta=\pi/4$, it is maximally entangled.

The final state is determined by the initial state and the $S$ matrix, 
\begin{align}
\label{finstate}
\vert \textup{fin}\rangle&=\cos\eta\vert p,\uparrow;q,\uparrow\rangle+\sin\eta ~e^{i\beta}\vert p,\downarrow;q,\downarrow\rangle
+i\sum_{\sigma_3,\sigma_4}\int_{\textbf{k}\neq\textbf{p}}\frac{2\pi\delta(E_{\textup{if}})}{2E_{\textbf{k}}2E_{\textbf{p}+\textbf{q}-\textbf{k}}}
\\ \notag
&\times\left[\cos\eta\mathcal{M}(k; \uparrow\uparrow,\sigma_3\sigma_4)+\sin\eta ~e^{i\beta}\mathcal{M}(k; \downarrow\downarrow,\sigma_3\sigma_4)\right]\vert k,\sigma_3;p+q-k,\sigma_4\rangle
\end{align}
where $\delta(E_{\textup{if}})=\delta(E_{\textup{fin}}-E_{\textup{ini}})$ and $\int_{\textbf{k}\neq\textbf{p}}\equiv\int d^3\textbf{k}/(2\pi)^3$. According to Ref.\cite{24}, delta functions are then regulated as 
\begin{align}
\label{deltafnc}
\delta^3_V(\textbf{p}-\textbf{p}^{\prime})=\frac{V}{(2\pi)^3}\delta_{\textbf{p},\textbf{p}^{\prime}},~~~~\delta_T(E-E^\prime)=\frac{1}{2\pi}\int_{-T/2}^{T/2}dt~e^{i(E-E^\prime)t}
\end{align}
where the setting of the entire scattering process is designated to occur in a large spacetime volume of duration $T$ and spatial volume $V$. Note that Eqs.$(\ref{deltafnc})$ imply $V=(2\pi)^3\delta_V^{(3)}(0)$ and $(2\pi)\delta_T(0)=T$. The factors $T$ and $V$ will be eliminated with a proper normalization.

From Eq.($\ref{finstate}$) we can evaluate the total density matrix of final state by $\rho_{AB}:=\vert \textup{fin}\rangle\langle\textup{fin}\vert$. The reduced density matrix $\rho^{\textup{(fin)}}_A$ is obtained by tracing out the degrees of freedom for particle $B$,  $\rho^{\textup{(fin)}}_A:=\mathcal{N}^{-1}~tr_B~\rho_{AB}$,  producing the following result
\begin{align}
\rho^{\textup{(fin)}}_A=&\frac{1}{\mathcal{N}}\biggr\{\cos^2\eta2E_{\textbf{q}}V\vert p,\uparrow\rangle
\langle p,\uparrow\vert+\sin^2\eta2E_{\textbf{q}}V\vert p,\downarrow\rangle
\langle p,\downarrow\vert
\\ \notag
+\lambda^2&\sum_{\sigma_3,\sigma^{\prime}_3}
\int_{\textbf{k}\neq\textbf{p}}
\frac{\{2\pi\delta(E_{\textup{if}})\}^2}{2E_{\textbf{k}}2E_{\textbf{p}+\textbf{q}-\textbf{k}}2E_{\textbf{k}}}
\mathcal{A}_{\sigma_3,\sigma^\prime_3}(\eta,\beta)
\vert k,\sigma_3\rangle\langle k,\sigma^{\prime}_3\vert\biggr\}
\end{align}
where $\mathcal{N}$ is the normalization factor fixed by $tr_A\rho^{\textup{(fin)}}_A=1$, 
\begin{align}
\mathcal{N}&=2E_{\textbf{q}}2E_{\textbf{p}}V^2+\lambda^2\int\limits_{\textbf{k}\neq\textbf{p}} 
\frac{\{2\pi\delta(E_{\textup{if}})\}^2V}{2E_{\textbf{k}}2E_{\textbf{p}+\textbf{q}-\textbf{k}}}
\mathcal{A}_{\sigma_3\sigma_3}(\eta,\beta)
\end{align}
and introducing a shorthand notation for the long expression
\begin{align}
\mathcal{A}_{\sigma_3\sigma^\prime_3}(\eta,\beta)=&\frac{1}{\lambda^2}\sum_{\sigma_4}(\cos\eta\mathcal{M}(k; \uparrow\uparrow,\sigma_3\sigma_4)+\sin\eta ~e^{i\beta}\mathcal{M}(k; \downarrow\downarrow,\sigma_3\sigma_4))
\\ \notag
&\times (\cos\eta\mathcal{M}^{\dagger}(k; \uparrow\uparrow,\sigma^\prime_3\sigma_4)+\sin\eta~ e^{-i\beta}\mathcal{M}^{\dagger}(k; \downarrow\downarrow,\sigma^\prime_3\sigma_4))
\end{align}
In the weak coupling, the reduced density matrix at order $\lambda^2$ can be written as
\begin{align}
\rho^{\textup{(fin)}}_A=\textup{diag}\left((1-\lambda^2 \mathcal{A})I_0,...,\lambda^2 \mathcal{A}_k,...\right)
\end{align}
where
\[I_0=\begin{pmatrix}
\cos^2\eta  &  0  \\
0  &  \sin^2\eta \\
\end{pmatrix}\]
\begin{align}
\label{AK}
\mathcal{A}=\int\limits_{\textbf{k}\neq\textbf{p}} 
\frac{\{2\pi\delta(E_{\textup{if}})\}^2}{2E_{\textbf{k}}2E_{\textbf{q}}2E_{\textbf{p}}2E_{\textbf{p}+\textbf{q}-\textbf{k}}V}\mathcal{A}_{\sigma_3\sigma_3}(\eta,\beta)
\end{align}
\[\mathcal{A}_k=\frac{\{2\pi\delta(E_{\textup{if}})\}^2}{2E_{\textbf{k}}2E_{\textbf{q}}2E_{\textbf{p}}2E_{\textbf{p}+\textbf{q}-\textbf{k}}V^2}
\begin{pmatrix}
\mathcal{A}_{11}(\eta,\beta)      &  \mathcal{A}_{12}(\eta,\beta) \\
\mathcal{A}_{21}(\eta,\beta)      &  \mathcal{A}_{22}(\eta,\beta) \\
\end{pmatrix} \]
Then the entanglement entropy between $A$ and $B$ in the final is $S^{\textup{(fin)}}_E=-tr\rho^{\textup{(fin)}}_A\log\rho^{\textup{(fin)}}_A$, and this is what we are going to study in the context of fermion-fermion scattering process. 

\subsection{Example: $e^+e^-\rightarrow\mu^+\mu^-$}
In the following, we will consider a simple reaction $e^+e^-\rightarrow\mu^+\mu^-$ in quantum electrodynamics (QED). The particle interaction during the scattering process induces change in the degree of entanglement between the particles from the incoming state to the outgoing state. We are interested in studying this variation of entanglement and shall proceed as follows.
\begin{figure}[htp]
  \centering
 \begin{varwidth}[t]{\textwidth} 
  \vspace{0pt}
  \includegraphics[scale=0.4]{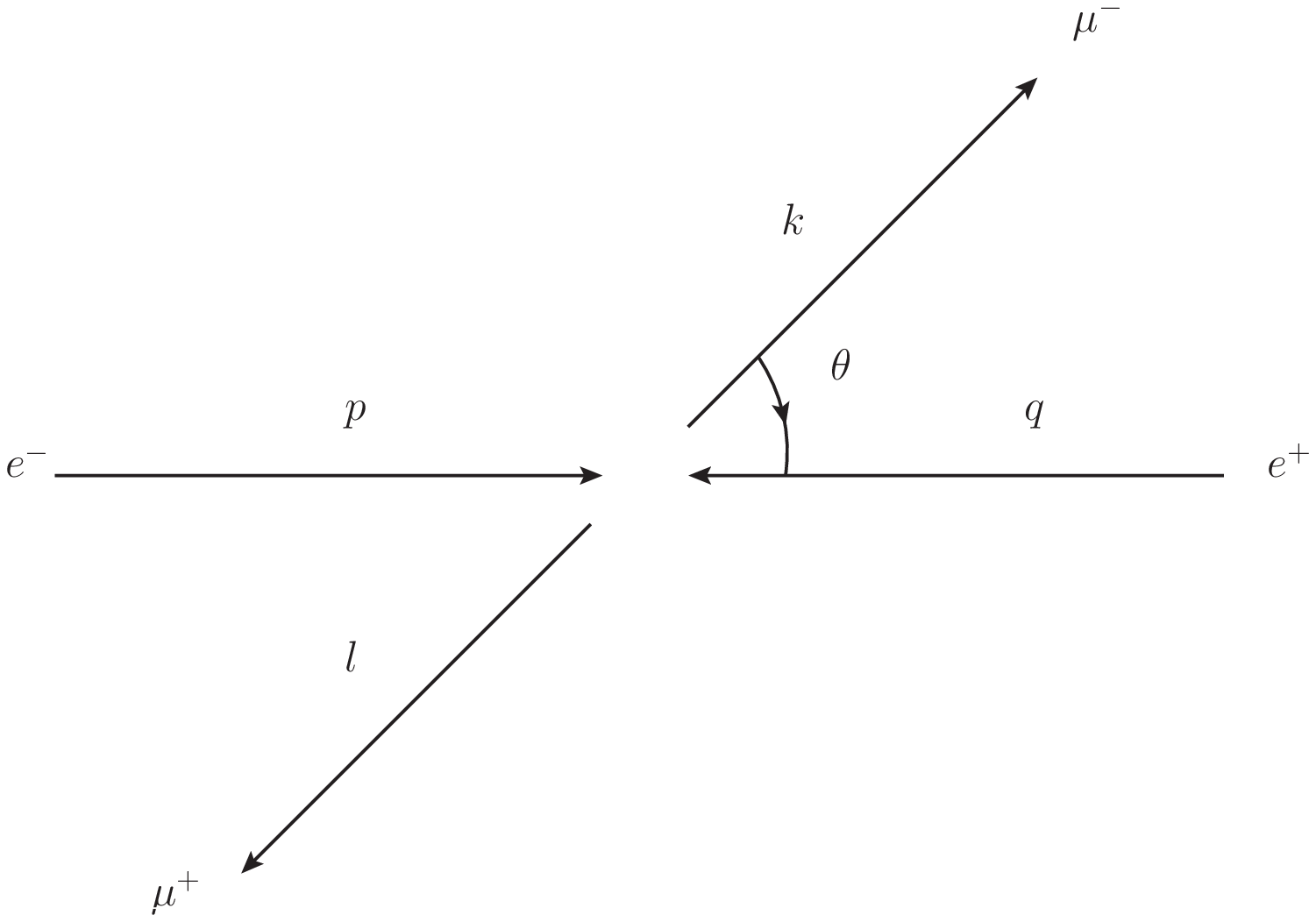}
  \end{varwidth}%
  \qquad
  \qquad
  \begin{varwidth}[t]{\textwidth}
  \vspace{0pt}
  \includegraphics[scale=0.5]{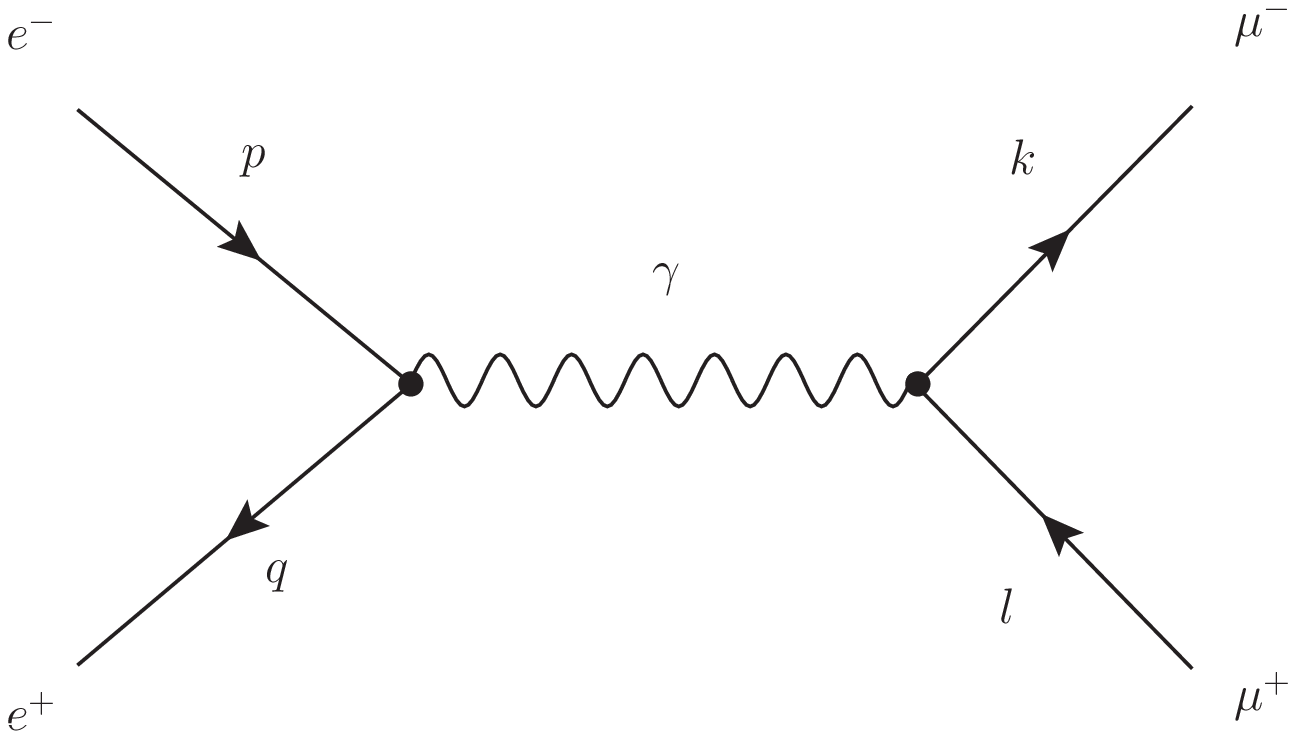}
  \end{varwidth}
 \caption{\textup{The QED annihilation process $e^+e^-\longrightarrow\mu^+\mu^-$ viewed in the center of mass frame and the corrsponding lowest order Feynman diagram}}
\end{figure}

The mass of particles is ignorable at high energy. Working in the center of mass frame, as the Fig. 1, the initial and final 4-momenta for $e^+e^-\rightarrow\mu^+\mu^-$ are
\begin{align}
p&=(E,0,0,E),~~~~q=(E,0,0,-E)
\\ \notag
k&=(E,E\sin\theta,0,E\cos\theta),~~~~l=(E,-E\sin\theta,0,-E\cos\theta)
\end{align}
The eigenvalues of matrix $A_k$ in Eq.($\ref{AK}$) can readily be obtained,
\begin{align}
a_{k1}=&\frac{1}{2}(3-2\cos\beta\sin2\eta\sin^2\theta+\cos2\theta-4\cos2\eta\cos\theta)
\\ \notag
a_{k2}=&\frac{1}{2}(3-2\cos\beta\sin2\eta\sin^2\theta+\cos2\theta+4\cos2\eta\cos\theta)
\end{align}
We then derive the entanglement entropy of the final state,
\begin{align}
S^{\textup{(fin)}}_E&=-(1-\lambda^2\mathcal{A})(\cos^2\eta\log\cos^2\eta+\sin^2\eta\log\sin^2\eta)+\lambda^2\mathcal{A}+\lambda^2\mathcal{A}\log(\frac{V^2}{T^2}\frac{16E^4}{\lambda^2})
\\ \notag
&-\frac{T}{V}\frac{\lambda^2}{128\pi^2 E^2 }\int d\Omega(a_{k1}\log(a_{k1})+a_{k2}\log(a_{k2}))
\end{align}
where
\begin{align}
\mathcal{A}=&\frac{T}{V}\frac{(1-\cos\beta\cos\eta\sin\eta)}{12\pi E^2}
\end{align}
For a given initial state, the change of entanglement entropy from initial to final state
\begin{align}
\label{varEE}
\triangle S_E&=\lambda^2\mathcal{A}(\cos^2\eta\log\cos^2\eta+\sin^2\eta\log\sin^2\eta)+\lambda^2\mathcal{A}+\lambda^2\mathcal{A}\log(\frac{V^2}{T^2}\frac{16E^4}{\lambda^2})
\\ \notag
&-\frac{T}{V}\frac{\lambda^2}{128\pi^2 E^2 }\int d\Omega(a_{k1}\log(a_{k1})+a_{k2}\log(a_{k2}))
\end{align}
In terms of the unpolarized total cross section for the scattering $e^+e^-\rightarrow\mu^+\mu^-$, $\sigma_{\textup{total}}=\frac{\lambda^2}{48\pi E^2}$, (Ref.\cite{28}) the term $\lambda^2\mathcal{A}$ in Eq.($\ref{varEE}$) becomes
\begin{align}
\lambda^2\mathcal{A}&=\frac{\sigma_tT}{V}f_1
,~~~~f_1=4(1-\cos\beta\cos\eta\sin\eta)
\end{align}
Thus, the variation of entanglement entropy would be
\begin{align}
\label{VEE}
\triangle S_E=\frac{\sigma_tT}{V}f_1\log\left[\frac{V^2}{T^2}\frac{16E^4}{\lambda^2}\right]+\frac{\sigma_tT}{V}g_1
\end{align}
where
\begin{align}
g_1&=f_1+f_2+f_3
\\ \notag
f_1&=4(1-\cos\beta\cos\eta\sin\eta)
\\ \notag
f_2&=4(1-\cos\beta\cos\eta\sin\eta)(\cos^2\eta\log\cos^2\eta+\sin^2\eta\log\sin^2\eta)
\\ \notag
f_3&=-\frac{3}{4}\int_{0}^{\pi}\sin[\theta](a_{k1}\log a_{k1}+a_{k2}\log a_{k2})
\end{align}
Note that the variation of entanglement entropy for the  scattering process of two scalar particles in Refs.\cite{21,22} correspond to the first term in our result Eq.($\ref{VEE}$). The extra term,  the second term comes from the entanglement of spin-spin and spin-momentum, which reflects the unique effect in fermionic system.

The variation of entanglement entropy between particle $A$ and $B$ in the scattering process $e^+e^-\rightarrow\mu^+\mu^-$ is plotted in Fig. 2.

\begin{figure}[htp]
  \centering
\begin{varwidth}[htp]{\textwidth} 
  \vspace{0pt}
  \includegraphics[scale=0.8]{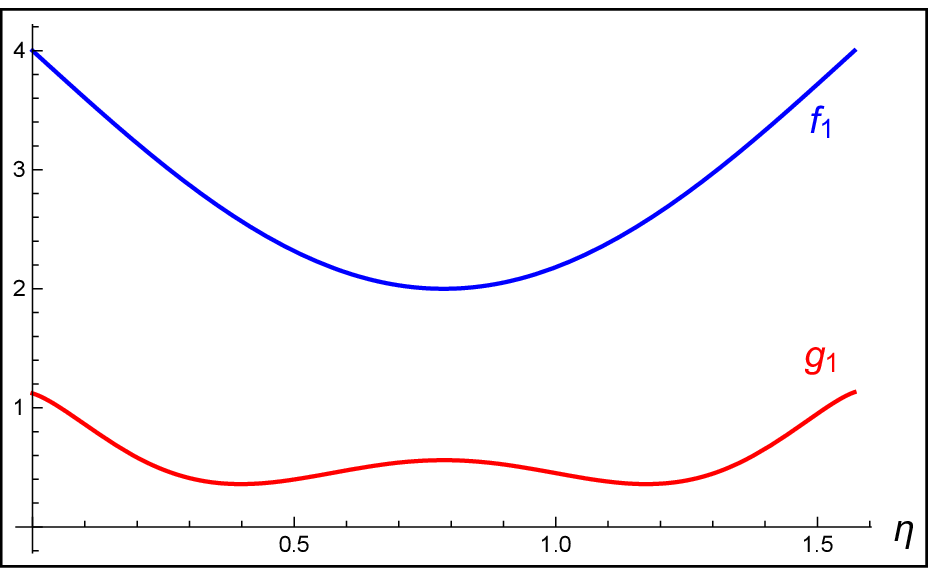}
 \end{varwidth}%
  \qquad
  \qquad
  \begin{varwidth}[htp]{\textwidth}
  \vspace{0pt}
  \includegraphics[scale=0.8]{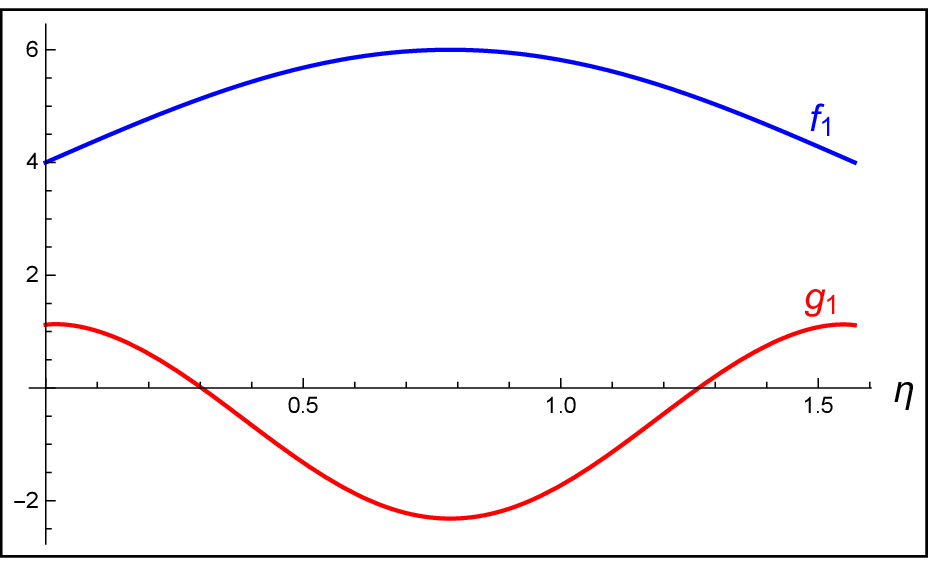}
  \end{varwidth}
 \caption{\textup{The variation of entanglement entropy $\triangle S_E$ as a function of entanglement parameter $\eta$ of the initial state. $\triangle S_E$ is proportional to the bottom curve $g_1$ plus the top curve $f_1$ multiplied by $\log[\frac{V^2}{T^2}\frac{16E^4}{\lambda^2}]$. Left figure for $\beta=0$. Right figure for $\beta=\pi$.}}
\end{figure}

From Eq.($\ref{VEE}$) we find that the variation of entanglement entropy is proportional to the quantity, $\sigma_tT/V$, where $\sigma_t$ is the unpolarizd total cross section for the above scattering process at the first order of perturbation theory. Spatial volume $V$ and duration $T$ denote  the spacetime volume encompassing the entire scattering process; they originate from the inner product of the single-mode state whose norm possesses a delta-functional divergence. In other words, ignoring the unphysical infinite factors $V$ and $T$, the variation of entanglement entropy in a scattering process is proportional to $\sigma_t$, the cross section. 
 In fact, the cross section is the effective area of a chunk taken out of one beam, by each particle in the other beam \cite{28}. So, the variation of entanglement entropy in a scattering process is shown to be proportional to the area quantity. We know that, for a ground state of a quantum many-body system, the entropy of the reduced state of a subregion often merely grows with the boundary area of the subregion, and not with its volume \cite{29,30,31}. Such ``area laws'' for the entanglement entropy emerge in several seemingly unrelated fields, in the context of black hole physics, quantum information science, and quantum many-body physics. Our results also seem to reflect the basic characteristics of entanglement entropy, area laws.   

\section{Mutual information between different degrees of freedom}

\subsection{Mutual information between spin and momentum}

In the above calculation, the initial state and final state can be regarded as generated by the basis of an asymptotically free Hamiltonian, and then their total Hilbert space can be divided into $\mathcal{H}_{\textup{tot}}=\mathcal{H}_A\otimes\mathcal{H}_B$. For a subsystem, say A, choosing spin state as the complete basis, its Hilbert space $\mathcal{H}_A$ can be further divided into $\mathcal{H}_{p_{A}}\otimes\mathcal{H}_{s_A}$, the spin and momentum degrees of freedom of subsystem $A$.
Then the total Hilbert space for the initial state and final state can be decomposed into $\mathcal{H}_{p_A}\otimes\mathcal{H}_{s_A}\otimes\mathcal{H}_{B}$. The mutual information between spin and momentum degrees of freedom for subsystem $A$,
\begin{align}
I(p_A,s_A)=S(p_A)+S(s_A)-S(p_A\cup s_A)
\end{align}
where $S(X)$ is the von Neumann entropy of the reduced density matrix of subsystem $X$. Mutual information is always greater than or equal to zero, with equality if and only if the density matrix for the subsystem $A$ is the tensor product of the reduced density matrices for subsysterms $p_A$ and $s_A$ for the initial state.

In the following, we calculate the mutual information between the spin and momentum for particle A in the scattering process $e^+e^-\rightarrow\mu^+\mu^-$. We use the same initial spin state parametrization  as ($\ref{ini})$
\begin{align}
\vert\textup{ini}\rangle=\cos\eta\vert p,\uparrow; q,\uparrow\rangle+\sin\eta~ e^{i\beta}\vert p,\downarrow; q,\downarrow\rangle
\end{align}
This gives zero value initial mutual information between spin and momentum degrees of freedom for particle $A$, $I^{\textup{(ini)}}=0$. As was explained in previous section, the final state and  reduced density matrix are determined by the initial state and $S$ matrix. In the analogous calculation, the mutual information between the spin and momentum of particle $A$ can be obtained.

 The momentum reduced density matrix for particle $A$ reads
 \begin{align}
\rho^{\textup{(fin)}}_{Ap}=\textup{diag}\left((1-\lambda^2 \mathcal{A}),...,\lambda^2B_k,...\right)
\end{align}
where
\begin{align}
\mathcal{A}&=\int\limits_{\textbf{k}\neq\textbf{p}}
\frac{\{2\pi\delta(E_{\textup{if}})\}^2}{2E_{\textbf{k}}2E_{\textbf{p}}2E_{\textbf{q}}2E_{\textbf{p}+\textbf{q}-\textbf{k}}V}\mathcal{A}_{\sigma_3\sigma_3}(\eta,\beta)
\\ \notag
B_k&=\frac{\{2\pi\delta(E_{\textup{if}})\}^2}{2E_{\textbf{k}}2E_{\textbf{p}}2E_{\textbf{q}}2E_{\textbf{p}+\textbf{q}-\textbf{k}}V^2}b_{k},~~b_k=\mathcal{A}_{\sigma_3\sigma_3}(\eta,\beta)
\end{align}
The corresponding  von Neumann entropy is
\begin{align}
 S^{\textup{(fin)}}_{p_A}&=\lambda^2\mathcal{A}+\lambda^2\mathcal{A}\log(\frac{16E^4V^2}{\lambda^2T^2})
 -\lambda^2\frac{T}{64\pi E^2 V}\int_{0}^{\pi}d\theta\sin\theta b_k\log b_k
 \\ \notag
 &=\frac{\sigma_tT}{V}f_1(1+\log\frac{V^2}{T^2}\frac{16E^4}{\lambda^2})+\frac{\sigma_tT}{V}f_4+\mathcal{O}(\lambda^4)
\end{align}
with $f_4=-\frac{3}{4}\int_{0}^{\pi}d\theta\sin\theta b_k\log b_k$.

Similarly, we have the spin reduced density matrix for particle $A$,
\[\rho^{\textup{(fin)}}_{s_A}=\begin{pmatrix}
S_{11} & S_{12} \\
S_{21} &   S_{22}  \\
\end{pmatrix} \]
with
\begin{align}
S_{11}&=(1-\lambda^2\mathcal{A})\cos^2\eta+\lambda^2C_{11}
\\ \notag
S_{12}&=\lambda^2C_{12}
,~~~~S_{21}=\lambda^2C_{21}
\\ \notag
S_{22}&=(1-\lambda^2\mathcal{A})\sin^2\eta+\lambda^2C_{22}
\\ \notag
C_{\sigma_3\sigma^\prime_3}&=\int\limits_{\textbf{k}\neq\textbf{p}} 
\frac{\{2\pi\delta(E_{\textup{if}})\}^2}{2E_{\textbf{k}}2E_{\textbf{p}}2E_{\textbf{q}}2E_{\textbf{p}+\textbf{q}-\textbf{k}}V}\mathcal{A}_{\sigma_3\sigma^\prime_3}(a,b)
\end{align}
And the eigenvalues of the density matrix $\rho^{\text{(fin)}}_{s_A}$ are found,
\begin{align}
c1&=\frac{1}{2}(1-\cos2\eta)+\lambda^2\frac{T}{V}\frac{1}{48\pi E^2}(\cos2\eta-\cos\beta\cos2\eta\sin2\eta))+\mathcal{O}(\lambda^4)
\\ \notag
c2&=\frac{1}{2}(1+\cos2\eta)-\lambda^2\frac{T}{V}\frac{1}{48\pi E^2}(\cos2\eta-\cos\beta\cos2\eta\sin2\eta))+\mathcal{O}(\lambda^4)
\end{align}
The corresponding von Neumann entropy 
\begin{align}
S^{\textup{(fin)}}_{s_A}&=-c1\log c1-c2\log c2
\\ \notag
&=\frac{\sigma_tT}{V}f_5+\cos^2\eta\log\cos^3\eta+\sin^2\eta\log\sin^2\eta+\mathcal{O}(\lambda^4)
\end{align}
where $f_5=(\log\cos^2\eta-\log\sin^2\eta)(\cos2\eta-\cos\beta\cos2\eta\sin2\eta)$.

For the given initial state, we thus obtain the variation of mutual information between  spin and momentum degrees of freedom for particle $A$ 
\begin{align}
\label{spmutual}
\triangle I(p_A,s_A)=\triangle S(p_A)+\triangle S(s_A)-\triangle S(A)=\frac{\sigma_tT}{V}g_2
\end{align}
where $g_2=f_4+f_5-f_2-f_3$, with functions $f_i$ given in previous sections. The variation of mutual information between spin and momentum for particle $A$ in the scattering process $e^+e^-\rightarrow\mu^+\mu^-$ is plotted in Fig. 3.

\begin{figure}[htp]
  \centering
 \begin{varwidth}[t]{\textwidth} 
  \vspace{0pt}
  \includegraphics[scale=0.7]{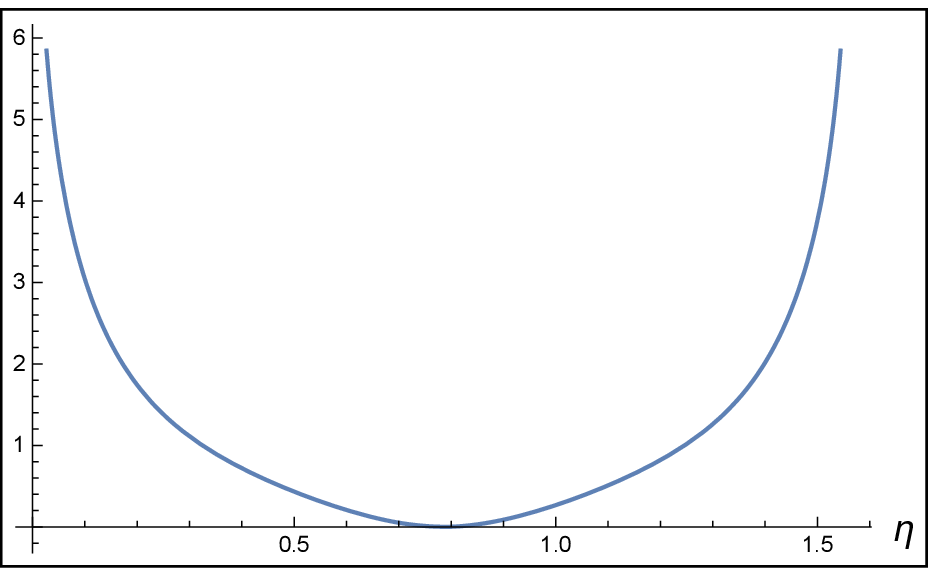}
  \end{varwidth}%
  \qquad
  \qquad
  \begin{varwidth}[t]{\textwidth}
  \vspace{0pt}
  \includegraphics[scale=0.7]{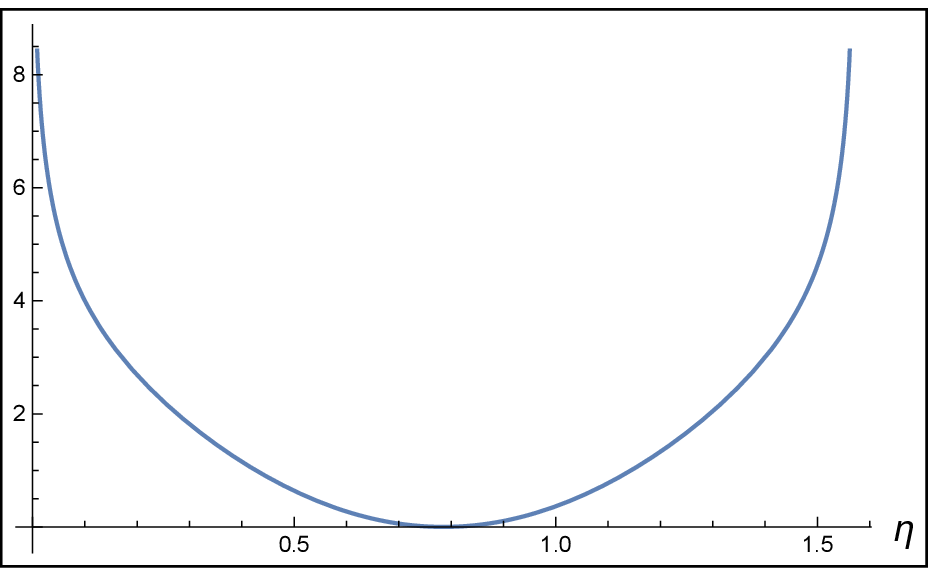}
  \end{varwidth}
 \caption{\textup{The variation of mutual information $\triangle I(p_A,s_A)$ as a function of entanglement parameter $\eta$. $\triangle I(p_A,s_A)$ is proportional to function $g_2(\eta)$. Left figure for the initial state with parameter $\beta=0$. Right figure for the initial state with parameter $\beta=\pi$.}}
\end{figure}

As explained above, the mutual information  vanishes for the initial state with arbitrarily valued entanglement parameters $\eta$  ($0\leq\eta\leq\pi/2$) and $\beta$ ($-\pi/2\leq\beta\leq3/2\pi$), that is,  $I^{\textup{(ini)}}=0$. From Eq.($\ref{spmutual}$) and Fig. 3, for and exclusively for a maximally entangled initial state between particles $A$ and $B$ ($\eta=\pi/4$), the mutual information for the final state is nearly zero. The zero value of mutual information means there is no entanglement between the spin and momentum degrees of freedom within particle $A$. Generally,  interactions cause change in the degree of entanglement between subsystems, but it is interesting to see that the degree of entanglement between the spin and momentum within  particle $A$ in the scattering process does not change when the initial state is  maximally entangled.    
 Then we conclude that the increasing degree of entanglement between two particles in the initial state restricts the generation of entanglement between the spin and momentum degrees of freedom within one particle in the final state. 

\subsection{Mutual information between helicity and momentum}
For one fermion, both helicity states and spin states  can be employed as the complete basis of Hilbert space. Now we choose the helicity states as the basis of two-particle Fock space for the initial state and final state. Then the total Hilbert space can be written as $\mathcal{H}=\mathcal{H}_{p_A}\otimes\mathcal{H}_{h_A}\otimes\mathcal{H}_B$. Furthermore, helicity states $\vert p,\lambda\rangle$ and spin states $\vert p,\sigma\rangle$ are related by the unitary transformation, Ref.\cite{27},
\begin{align}
\vert p,\lambda\rangle=D_{\sigma\lambda}[R(p)]\vert p,\sigma\rangle
\end{align}
where $R(p)$ is the rotation that carries the $z$ axis into the direction $\textbf{p}$, and $D$ is the spin 1/2 irreducible unitary representation of Lorentz group,
\[ D[R(p)]=\begin{pmatrix}
e^{-i\frac{\phi}{2}} & 0  \\
0 & e^{i\frac{\phi}{2}}
\end{pmatrix}
\begin{pmatrix}
\cos\frac{\theta}{2} &  -\sin\frac{\theta}{2} \\
\sin\frac{\theta}{2}  &  \cos\frac{\theta}{2}
\end{pmatrix}  \]
with $\hat{\textbf{p}}=(\sin\theta\cos\phi,\sin\theta\sin\phi,\cos\theta)$, and labeling helicity $\lambda= 1/2,-1/2$ as $1,2$, respectively.

From the description of the initial entangled state in the spin-based Hilbert space, passing to its description in the helicity-based Hilbert space, we have
\begin{align}
\vert \textup{ini}\rangle&=\cos\eta\vert p,\uparrow;q,\uparrow\rangle+\sin\eta e^{i\beta}\vert p,\downarrow;q,\downarrow\rangle
\\ \notag
&=\sum_{\lambda_1,\lambda_2}(\cos\eta D^{-1}_{\lambda_11}[p]D^{-1}_{\lambda_21}[q]+\sin\eta~e^{i\beta} D^{-1}_{\lambda_12}[p]D^{-1}_{\lambda_22}[q])\vert p,\lambda_1; q,\lambda_2\rangle
\end{align}
We can readily obtain the final state,
\begin{align}
\vert \textup{fin}\rangle&=
\sum_{\lambda_1,\lambda_2}(\cos\eta D^{-1}_{\lambda_11}[p]D^{-1}_{\lambda_21}[q]+\sin\eta~e^{i\beta} D^{-1}_{\lambda_12}[p]D^{-1}_{\lambda_22}[q])\vert p,\lambda_1; q,\lambda_2\rangle
\\ \notag
&+i\sum_{\lambda_3,\lambda_4}\int_{\textbf{k}\neq\textbf{p}}
\frac{2\pi\delta(E)}{2E_{\textbf{k}}2E_{\textbf{p}+\textbf{q}-\textbf{k}}}
\vert \textbf{k},\lambda_3;\textbf{p}+\textbf{q}-\textbf{k},\lambda_4\rangle
\\ \notag
&\times D^{-1}_{\lambda_3\sigma_3}[k]D^{-1}_{\lambda_4\sigma_4}[l](\cos\eta\mathcal{M}(k; \uparrow\uparrow,\sigma_3\sigma_4)+\sin\eta~e^{i\beta}\mathcal{M}(k; \downarrow\downarrow,\sigma_3\sigma_4))
\end{align}
The reduced density matrix $\rho_A^{(h, \textup{fin})}$ can be written as
\begin{align}
\rho_A^{(h, \textup{fin})}&=\frac{1}{\mathcal{N}^{\prime}}
\biggr\{\sum_{\lambda_1,\lambda^\prime_1}D_{\lambda_1\lambda^\prime_1}(\eta,\beta)
2E_{\textbf{q}}V\vert p,\lambda_1\rangle\langle p,\lambda^\prime_1\vert
\\ \notag
&+\lambda^2\sum_{\lambda_3,\lambda^\prime_3}
\int_{\textbf{k}\neq\textbf{p}} 
\frac{\{2\pi\delta(E)\}^2}{2E_{\textbf{k}}2E_{\textbf{p}+\textbf{q}-\textbf{k}}2E_{\textbf{k}}}\mathcal{A}^\prime_{\lambda_3,\lambda^\prime_3}(\eta,\beta)\vert k,\lambda_3\rangle\langle k,\lambda^\prime_3\vert
\biggr\}
\end{align}
where
\begin{align}
\notag
D_{\lambda_1\lambda^\prime_1}(\eta,\beta)&=\sum_{\lambda_2}
\left(\cos\eta D^{-1}_{\lambda_11}[p]D^{-1}_{\lambda_21}[q]
+\sin\eta~e^{i\beta}D^{-1}_{\lambda_12}[p]D^{-1}_{\lambda_22}[q]\right)
\\ \notag
&\times\left(\cos\eta~D_{1\lambda^\prime_1}[p]D_{1\lambda_2}[q]+\sin\eta~e^{-i\beta}D_{2\lambda^\prime_1}[p]D_{2\lambda_2}[q]\right)
\\ 
\mathcal{A}^\prime_{\lambda_3\lambda^\prime_3}(\eta,\beta)&=\frac{1}{\lambda^2}D^{-1}_{\lambda_3\sigma_3}[k]D^{-1}_{\lambda_4\sigma_4}[l](\cos\eta~M(k; \uparrow\uparrow,\sigma_3\sigma_4+\sin\eta~e^{i\beta}M(k; \downarrow\downarrow,\sigma_3\sigma_4)
\\ \notag
&\times(\cos\eta~M^{\dagger}(k; \uparrow\uparrow,\sigma^\prime_3\sigma^\prime_4)+\sin\eta~e^{-i\beta}M^{\dagger}(k; \downarrow\downarrow,\sigma^\prime_3\sigma^\prime_4))D_{\sigma^\prime_3\lambda^\prime_3}[k]D_{\sigma^\prime_4\lambda_4}[l]
\end{align}
and normalization factor  $\mathcal{N}^{(h)}$ is fixed by $tr_A\rho_A^{(h, \textup{fin})}=1$,
\begin{align}
\mathcal{N}^{(h)}&=D_{\lambda_1\lambda_1}(\eta,\beta)2E_{\textbf{p}}2E_{\textbf{q}}V^2+\lambda^2\int\limits_{\textbf{k}\neq\textbf{p}}
\frac{\{2\pi\delta(E)\}^2V}{2E_{\textbf{k}}2E_{\textbf{p}+\textbf{q}-\textbf{k}}}
\mathcal{A}^\prime_{\lambda_3\lambda_3}(\eta,\beta)
\end{align}

For the scattering process $e^+e^-\rightarrow\mu^+\mu^-$, we consider that the helicity states and spin states are observed within the same inertial reference frame. Straightforward calculations show that the reduced density matrix $\rho_A^{(h, \textup{fin})}$ in helicity representation and $\rho^{\textup{(fin)}}_{A}$ in spin representation have the same eigenvalues. This is a natural conclusion, since the entanglement entropy between subsystems A and B is irrelevant of the basis of Hilbert space,  $S^{\textup{(fin)}}_A=S_A^{(h,\textup{fin})}$. Similarly, we find that the von Neumann entropy $S^{(h,\textup{fin)}}_{p_A}$ of the momentum reduced density matrix in helicity representation is equal to the von Neumann entropy $S^{\textup{(fin)}}_{p_A}$ in spin representation, $S^{(h,\textup{fin)}}_{p_A}=S^{\textup{(fin)}}_{p_A}$.

To proceed with the analysis, noting that the helicity  reduced density matrix of particle $A$
\[\rho^{\textup{(fin)}}_{h_A}=\begin{pmatrix}
h_{11} & h_{12} \\
h_{21} &  h_{22} \\
\end{pmatrix} \]
with
\begin{align}
h_{11}&=(1-\lambda^2\mathcal{A}^\prime)D_{11}(\eta,\beta)+\lambda^2\int\limits_{\textbf{k}\neq\textbf{p}} 
\frac{\{2\pi\delta(E)\}^2}{2E_{\textbf{k}}2E_{\textbf{p}}2E_{\textbf{q}}2E_{\textbf{p}+\textbf{q}-\textbf{k}}V}\mathcal{A}^\prime_{11}(\eta,\beta)
\\ \notag
h_{12}&=(1-\lambda^2\mathcal{A}^\prime)D_{12}(\eta,\beta)+\lambda^2\int\limits_{\textbf{k}\neq\textbf{p}} 
\frac{\{2\pi\delta(E)\}^2}{2E_{\textbf{k}}2E_{\textbf{p}}2E_{\textbf{q}}2E_{\textbf{p}+\textbf{q}-\textbf{k}}V}\mathcal{A}^\prime_{12}(\eta,\beta)
\\ \notag
h_{21}&=(1-\lambda^2\mathcal{A}^\prime)D_{21}(\eta,\beta)+\lambda^2\int\limits_{\textbf{k}\neq\textbf{p}} 
\frac{\{2\pi\delta(E)\}^2}{2E_{\textbf{k}}2E_{\textbf{p}}2E_{\textbf{q}}2E_{\textbf{p}+\textbf{q}-\textbf{k}}V}\mathcal{A}^\prime_{21}(\eta,\beta)
\\ \notag
h_{22}&=(1-\lambda^2\mathcal{A}^\prime)D_{22}(\eta,\beta)+\lambda^2\int\limits_{\textbf{k}\neq\textbf{p}} 
\frac{\{2\pi\delta(E)\}^2}{2E_{\textbf{k}}2E_{\textbf{p}}2E_{\textbf{q}}2E_{\textbf{p}+\textbf{q}-\textbf{k}}V}\mathcal{A}^\prime_{22}(\eta,\beta)
\end{align}
has the following roots:
\begin{align}
h_1=\frac{1}{2}(1-\cos2\eta)+\lambda^2\frac{T}{48\pi E^2V}(2\cos2\eta-\cos\beta\sin2\eta\cos2\eta)
\\ \notag
h_2=\frac{1}{2}(1+\cos2\eta)-\lambda^2\frac{T}{48\pi E^2V}(2\cos2\eta-\cos\beta\sin2\eta\cos2\eta)
\end{align}
Hence, the variation of helicity entanglement entropy follows
\begin{align}
\triangle S(h_A)=\frac{\lambda^2T}{48\pi E^2V}(\log\cos^2\eta-\log\sin^2\eta)(2\cos2\eta-\cos\beta\sin2\eta\cos2\eta)=\frac{\sigma_tT}{V}f_6
\end{align}
with $f_6=(\log\cos^2\eta-\log\sin^2\eta)(2\cos2\eta-\cos\beta\sin2\eta\cos2\eta)$.

At last, we obtain the variation of mutual information between helicity and momentum degrees of freedom of particle $A$,
\begin{align}
\triangle I(p_A,h_A)=\triangle S(p_A)+\triangle S(h_A)-\triangle S(A)=\frac{\sigma_tT}{V}g_3
\end{align}
where $g_3=f_4+f_6-f_2-f_3$. We compare the corresponding variations of mutual information of the helicity-momentum and spin-momentum entanglements in Fig. 4.

\begin{figure}[htp]
  \centering
 \begin{varwidth}[t]{\textwidth} 
  \vspace{0pt}
  \includegraphics[scale=0.8]{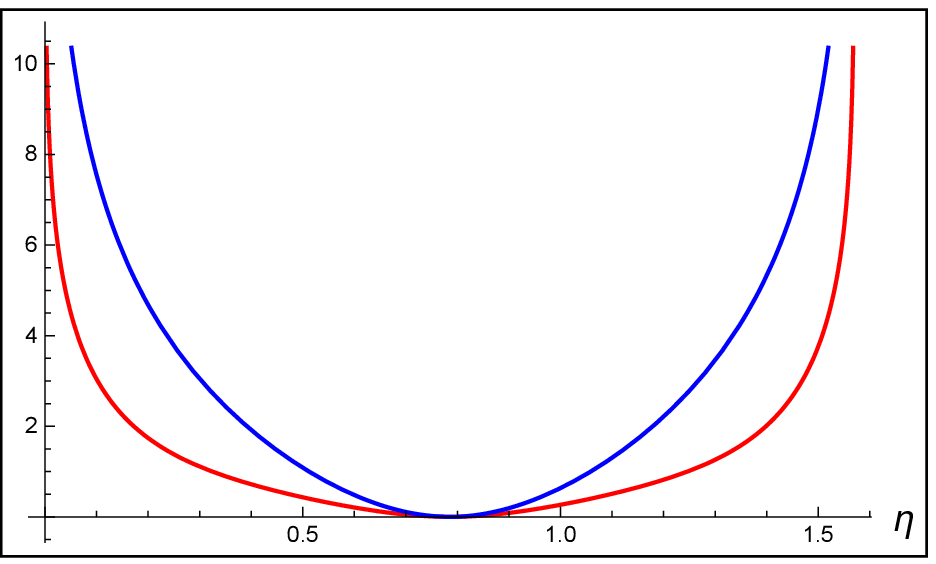}
  \end{varwidth}%
  \qquad
  \qquad
  \begin{varwidth}[t]{\textwidth}
  \vspace{0pt}
  \includegraphics[scale=0.8]{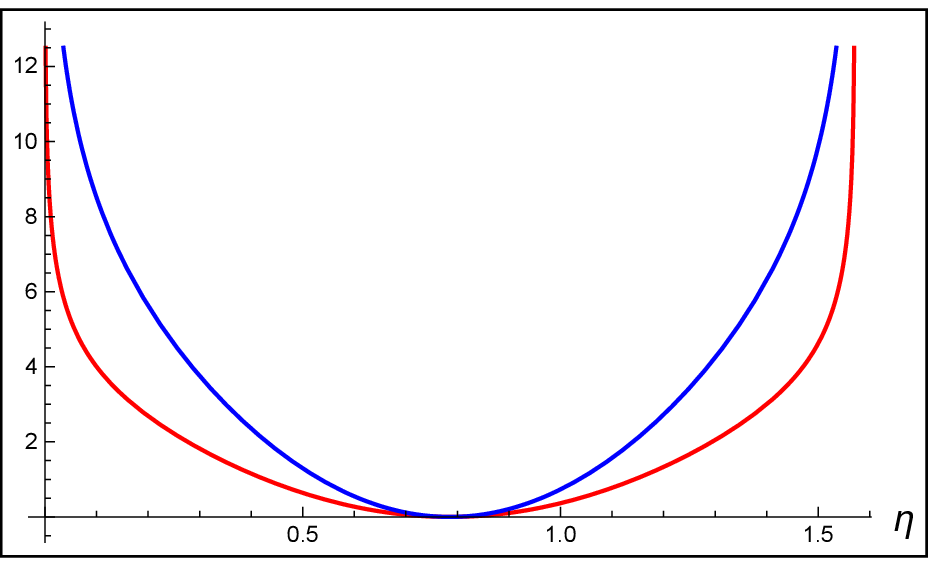}
  \end{varwidth}
 \caption{\textup{For the scattering process $e^+e^-\rightarrow\mu^+\mu^-$, the variation of mutual information $\triangle I$ in helicity representation (red curve) vs spin representation (blue curve) with the same inertial reference frame as a function of entanglement parameter $\eta$ form $0$ to $\pi/2$. Left figure for the initial state with parameter $\beta=0$. Right figure for the initial state with parameter $\beta=\pi$}.}
\end{figure}

As mentioned above, our starting setting is $I^{\textup{(ini)}}=0$. From Fig. 4, the final state mutual information $I^{\textup{(fin)}}(p_A, s_A)$ between spin and momentum is different from that between helicity and momentum. Both of them approach zero, if the initial state is a maximally entangled state ($\eta=\pi/4$). We conclude that the generation of the final state entanglement of both spin-momentum and helicity-momentum correlations within one particle are restricted by the increasing degree of the initial state entanglement between two particles. And for a general initial entanglement state of two particles, the degree of the final state  entanglement between spin and momentum within one particle is always stronger than  that for helicity-momentum. It indicates that the helicity and spin should have significantly distinct properties of entanglement perceived by an inertial observer, confirming a similar observation in literature Refs.\cite{26,27}.

\section{Conclusion}

For an elastic scattering process of two fermionic particles, we performed a detailed study of the variation of entanglement entropy from an initial state to the final state. With the QED scattering process $e^+e^-\rightarrow\mu^+\mu^-$ being employed as an example of the case study, the variation of entanglement entropy was computed  with respect to different entanglement of the initial state. The mathematical expression of our result contains two terms of contribution. The first term corresponds to the variation of entanglement for the scalar scattering particles, and the extra term reflects the unique effect in the  fermionic system. This variation is found to be proportional to $\sigma_tT/V$, with $\sigma_t$ being the cross section, and $T$, $V$ ignorable unphysical infinity artifacts  coming from the delta function regularization. Therefore, coincident with the basic characteristic of entanglement entropy $-$ the area laws, as being expected, this result was obtained in the context of the  fermionic system and allows for an extension to general field systems.

What might be more interesting is the behavior of the entanglement between spin (or helicity) and momentum degrees of freedom within one particle for a two-particle composite system in a scattering process. For the aforementioned QED scattering process, the relevant variations of mutual information for both these cases were calculated in the same inertial reference frame. For a maximally entangled initial state, it was found that the scattering between the particles does not affect the degree of either the spin-momentum or the helicity-momentum entanglement of one particle in the final state. Furthermore, we found that the higher degree of entanglement between two ingoing particles would restrict the generation of entanglement between the spin/helicity and momentum of one of the outgoing particles. The entanglement between spin and momentum within one outgoing particle is shown always stronger than that for helicity-momentum for a general initial entanglement state,  implying significantly distinct properties of entanglement for the helicity and spin perceived by an inertial observer.

It might be interesting to study the entanglement between the spin degrees of freedom for two particles with $I(s_A, s_B)=S(s_A)+S(s_B)-S(s_{A\cup B})$, for which the total Hilbert space can be divided into $\mathcal{H}_{p_A}\otimes\mathcal{H}_{s_A}\otimes\mathcal{H}_{p_B}\otimes\mathcal{H}_{s_B}$. It can be found that the roots of the reduced density matrix of spin-subsystem $s_{A\cup B}$ are complex values. Or to consider a different scattering process, such as $e^+e^-\longrightarrow e^+e^-$, the amplitude would contain a sum of $s$-channle and $t$-channel Feynman diagrams, for which the initial state needs be chosen as $\vert\textup{ini}\rangle=a~\vert p,\uparrow; q,\uparrow\rangle+b~\vert p,\downarrow; q,\downarrow\rangle+c~\vert p,\uparrow; q,\downarrow\rangle+d~\vert p,\downarrow; q,\uparrow\rangle$, with $\vert a\vert^2+\vert b\vert^2+\vert c\vert^2+\vert d\vert^2=1$. Then the variation of entanglement entropy would become a function of multiple parameters, which is tricky to analyze at present but worthy of further study.

\paragraph{\zihao{4}Acknowledgment}
\paragraph{}
The authors would like to thank Dingfang Zeng and Xinfei Li for useful discussions. This work is supported by the National Natural Science
 Foundation of China (Grants No. 11275017 and No. 11173028).

\end{document}